# Catching Attention in Fiber Optics Class

R. Ya. Kezerashvili, L. Leng

New York City College of Technology, The City University of New York

## 1. Introduction

The past 10 years has witnessed the rapid growth of the Internet traffic fueled by World Wide Web applications. It is estimated that nearly half a billion people have Internet access and use it regularly. The World Wide Web already hosts over 2 billion web pages, and, according to the estimates, people upload more than 3.5 million new web pages every day. Reports have shown that data traffic flowing over the Internet approximately doubles every year [1]. The increase of the Internet traffic has resulted in enormous requirements for the bandwidth of information transport. Among a number of technologies, optical fiber is considered to be the best suited one for providing this bandwidth.

An optical fiber is a hair-thin strand of glass or plastic through which communication signals are transmitted from one location to another in the form of light generated by lasers or other signal sources. These signals are digital pulses or continuously modulated analog streams of light representing voice information, data information, video information, or any other type of information.

Optical fiber was originally invented in the early 1970s for long-distance voice communication. AT&T put the first commercial optical fiber telephone system into service in 1983 [2]. However, it was the rapidly rising demand for the Internet connectivity that drove the growth and development of fiber optic communications. Today optical fiber is gradually replacing the traditional transmission media (copper wire, coaxial cable, etc.) because it has larger information capacity, lower loss, less weight, and better immunity to electromagnetic interference. Further more, people are trying to replace electrical switches, regenerators, and buffers with their optical counterparts in order to gain higher bandwidth and faster speed.

Nowadays there are various optical networks deployed around the world, which use optical fibers to connect different locations and to carry data, voice, or video signals. These include backbone networks that provide country-to-country or city-to-city transport services, metropolitan area networks that link access networks with the backbone, and access networks that provide cable TV and fiber-to-the-home services. Due to these deployments we are enjoying cheaper telephone calls and high-speed Internet services. Every day new technologies and bandwidth-hungry applications (such as remote teaching, remote monitoring, video on demand, etc. [3]) are pushing optical fiber closer to our homes and offices. In addition to telecommunications, there are also many other applications for optical fiber that are simply not possible with metallic conductors. These include sensors, medical/surgical applications, subject illumination, image transport, military, automotive, and industrial applications [4].

Since optical fiber technology is growing to be part of our daily lives, the understanding of its applications will be critical knowledge for our society, especially for the younger generations. To better prepare the students for the fast-pacing information age, college education and training on fiber optics has apparently become high priority.

## 2. What is fiber optics?

According to American National Standard for Telecommunications, fiber optics is the branch of optical technology concerned with the transmission of light through fibers made of transparent materials such as glasses and plastics. We would like to point out that telecommunication applications of fiber optics mostly use flexible low-loss fibers made of pure silica glass, which will be the focus of this article. Current plastic fibers suffer from high loss, which presents both technical and economical challenge for their wide applications.

Light can be transmitted in an optical fiber with a loss as low as physically possible due to a blessed natural phenomenon: total internal reflection, which occurs when light propagates from a medium with a high index of refraction to one with a lower index of refraction [5]. In optical fiber, the core - the center region through which light is transmitted - has a slightly higher index of refraction than its surrounding ring, the cladding, as shown in Fig. 1. When total internal reflection occurs, the traveling light in the core hits the cladding and reflects back without losing any light to the cladding. The loss that the light experiences during propagation is due to scattering and absorption in the fiber and bending of the fiber.

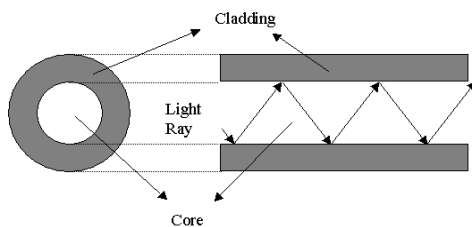

Fig. 1. Schematic illustration of the cross section of a typical optical fiber and light propagation in optical fiber

The idea of guiding light based on total internal reflection dates back to 1870, when John Tyndall, a British scientist, used a jet of water flowing from one container to another and a beam of light to demonstrate that light inside the water guide followed a specific path. The first all-glass fiber came about in the 1950s as an image-transmitting device called fiberscope. However, these early fibers experienced excessive loss and, as a result, the transmitting distances were severely limited. It was the advent of laser technology and low-loss glass fiber that enabled the establishment of the fiber optics industry in the 1970s [6]. Back then, military and commercial uses of optical fibers for telephone networks were the driving force for the progress of fiber optic communications. During the 1990s, the Internet started to embrace fiber optics, and its phenomenal growth put optical networking into the focus of new technologies.

A basic fiber optic system consists of three elements: a transmitter, which generates light signal using a laser or a light-emitting diode; an optical fiber cable, which carries the light to its destination; and a receiver, which accepts the light signal transmitted. On the transmitter side, information such as data, voice, and video, is first encoded into electrical signal and then converted into optical signal at the laser source. The optical signal propagates along the fiber until it reaches the receiver, where the optical signal is converted back into electrical signal. As the last step, the electrical signal is decoded into information in the form of data, voice, or video [7].

More sophisticated fiber optic systems may contain fiber amplifiers for compensating the transmission loss, intelligent switches for directing the light traffic in optical networks, signal regenerators for boosting the signal quality and cleaning the noise, etc.

### 3. How to get students interested?

As implied by its definition, fiber optics covers a broad range of knowledge and, therefore, is generally interdisciplinary. The goal of a fiber optics course is to teach students of science major the physical fundamentals of fiber optics and to stimulate interests that may potentially develop into the students' future careers. The objectives of the fiber optics lecture for non-science major students are to explain the basic principles, to present rapidly evolving fiber optic technologies that are supporting our modern society's daily communication activities, and to foresee the impact of fiber optics on various aspects of our life. Of course these objectives are applicable for science major students, too.

Fiber optics in general requires some preliminary knowledge of properties of light. Suggested topics include: nature of light, refractive index, Snell's law, total internal reflection, polarization and birefringence, dispersion, interference and interferometers, diffraction and diffraction gratings. Depending on the students targeted with the course, these topics can be either required as preliminary or taught at the beginning of the course. After students have the basic knowledge of geometric and wave optics, the fundamental theory of waveguide with the focus on optical fiber can be introduced. The other suggested topics, which can be included in the syllabus, are fiber structure and fabrication, laser physics, opto-electronics, etc. Like any other disciplines, fiber optics can be both challenging and boring, which makes it important to keep the students motivated and interested, especially during the lectures. We find that the most effective approach is to demonstrate some simple yet revealing fiber optics experiments right before the lectures. Often these demonstrations bring out the curiosity in the students and help them to focus on the following materials covered in the lecture. Such demonstrations can be also modified and included in the fiber optics laboratory projects. We find that students, especially of non-science major, can learn more easily and gain better understanding of fiber optics with the help of these demonstrations. In this section, we provide and describe a few examples of such experiments.

3.1. A basic fiber optic transmission system

In the very first few lectures, the teacher brings to the class a laser pointer, a small portable light source with fiber-coupled output, a screen, and a strand of fiber optics cable with connectors on both ends. These components can be purchased from fiber optics manufactures or vendors such as AC Photonics Inc. and Fiber Instrument Sales [8]. First, the teacher aims the laser pointer at the screen through the chalk dust (use an eraser to make the chalk dust) and shows that the light path is a straight line. This also visually helps the students to understand the concept of a light ray in geometric optics. Then the teacher sets up the demonstration as shown in Fig. 2, and shows that light can be coupled into a coiled fiber optic cable and still emerges at the other end. How is this possible? Can light be bent? What does the fiber cable do to the light so that it can be bent and follows the guide of the fiber cable? Naturally those questions arise in the students' minds. Now is the time for the teacher to point out that the demonstration is a simplified version of a real-life communication system, in which information is carried from one location to another through fiber optics. In fact, the phone calls they make, the downloaded files in their computers, and their favorite sports TV programs, have all traveled through some certain fiber cables. And the answers to the questions are exactly what the fiber optics course is about. The students are now informed that they have to know about a few physical aspects of light: reflection, refraction, and diffraction, before they can truly and fully understand the answer. At this time they realize that what they are about to learn is not just some formulas or figures in the textbook, but also real-life phenomena that are very close to them. Moreover, the level of knowledge absorption will not only affect their grades, but also might have an impact on their future careers.

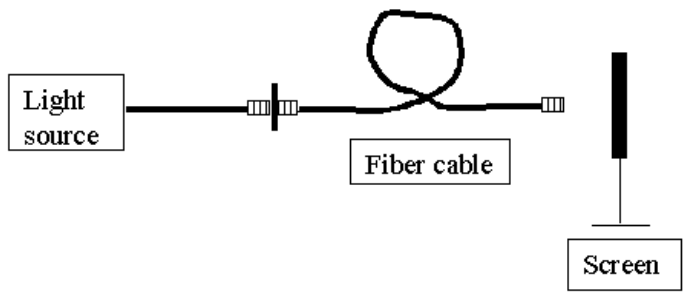

Fig. 2. A basic fiber optic transmission system

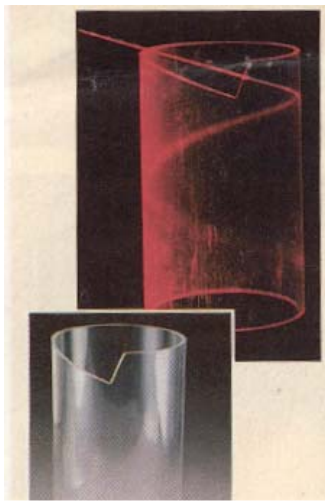

Fig. 3. Internal reflection tube (Photo courtesy of CENCO)

Corresponding to the lecture demonstration, the students can perform a no-prior-experience-needed experiment using an internal reflection tube [9] as a laboratory project. The internal reflection tube is a specially designed, clear acrylic tube that measures 7.5 cm in diameter and 12.5 cm in height, as shown in Fig. 3. In this experimental activity the students can clearly witness internal reflections when they shine a laser beam onto a polished step cut into the upper end of this tube. The beam's path is clearly visible as a spiral inside the tube wall - making for a fascinating demonstration of internal light reflections.

3.2. Water jet optics

This demonstration is similar to John Tyndall's experiment in 1870 [10] and essentially demonstrates the same principle: total internal reflection. As shown in Fig. 4, a light source (in our setup, a He-Ne laser, or diode laser) is positioned and adjusted so that its output light beam horizontally enters a water container and exits through a spout on the other side of the container. Now the teacher can seal the container with a cap and ask the students to pay attention to the screen. What do they see? A light spot, the students will answer. Then, the teacher removes the cap and asks the same question again. The students will notice that the light spot has disappeared from the screen. What happened to the light spot? Now it is a good time to encourage them to come up with an answer. When the container is sealed with a cap, no water flows out of the spout due to the air pressure. The light beam continues propagating in its original direction after leaving the container and reaches the screen creating a bright spot. When the cap is removed, water will flow through the spout naturally forming a water waveguide. Since the refractive index of water is larger than that of air, the condition for total internal reflection can be satisfied. As a result, the light beam exiting the spout goes through multiple total internal reflections within the water guide thus following the water flow instead of the horizontal path. If the frequency of the light beam is within the visible spectrum range of human's eyes, then one can actually see the light traveling within the water. In a dark room, the effect of this experiment is impressive. "One of the most beautiful and most curious experiments that one can perform in a course on optics", professor Daniel Colladon at the University of Geneva wrote about water jet experiment [6].

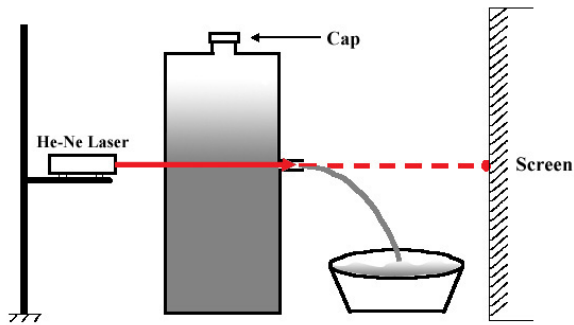

Fig. 4. Setup of water jet optics

This is an interesting experiment, the students may think, but what does it have to do with fiber optics? A question like that would make the teacher happy because now he/she can discuss fiber optics. He/she could start with "well, this demonstration shows exactly how fiber optics works. Water jet optics and fiber optics are based on the same principle: total internal reflection. If you understand this experiment, you've practically known the fundamentals of fiber optics. However, fiber optics is a much better waveguide than a water jet, thanks to numerous technology breakthroughs in the last fifty years. And you are about to explore all that accumulated knowledge in this course …"

For the corresponding laboratory project, a light pipe [9] can be used to observe total internal reflection. The students can use a crystal clear 0.80-m long acrylic rod bended into a spiral with a long tail. When a laser light is shined into one end of the rod it will emerge from the other with almost no attenuation.

3.3. Light wave propagation in a glass rod

The propagation of light wave in optical fiber is the main body of the fiber optics course. Students are required to understand the concepts and mechanisms of fiber loss, dispersion, birefringence, etc. Since the light wave is confined in the fiber, which is as thin as human hair, most of the physical phenomena cannot be visually perceived, which hinders students' learning process. We find it a very helpful demonstration to show a blown-up version of fiber optics: light propagation in a glass rod. This seemingly simple experiment shown in Fig. 5 is actually very rich in content.

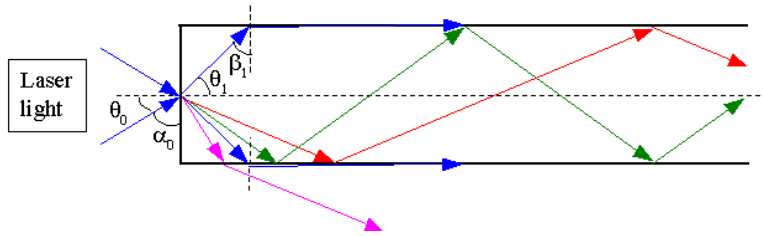

Fig. 5. Light wave propagation in a glass rod

First, it can help the students to understand how light waves are confined in fiber optics through the mechanism of total internal reflection. When a laser light is injected into the glass rod at an incident angle α, depending on the value of α, the propagation of the light varies significantly. A particularly interesting case is represented by the blue color in Fig. 5, where the laser light is adjusted to enter the glass rod with a refractive angle of $\theta_1$ and to subsequently meet the interface between the glass and the air at an angle of $\beta_1$, so that the refracted light travels in parallel with the glass rod. Why does this happen? What does this phenomenon imply? The teacher can get the students actively involved in the thinking process by presenting those questions. He/she can also explain that it occurs because the refractive index of the glass is higher than that of the air. After this step, we know that, based on geometric optics, any light that enters the glass rod with an incident angle of α that is greater than $\alpha_0$ will reflect back into the glass rod without any loss and is, therefore, confined in the glass rod. The green and red arrowed lines plotted in Fig. 5 are examples of such light rays. On the contrary, an incident angle less than $\alpha_0$ results in the loss of the light, as illustrated by the purple lines. Now it is the time for the teacher to introduce the concepts of "acceptance angle" and "numerical aperture (NA)", which are commonly used terms for characterizing fiber optics. The acceptance angle is the maximum angle from the fiber axis at which the light may enter the fiber so that it will reflect and travel correctly in the fiber. It is shown in Fig. 5 as the angle $\theta_0$. The sine of the acceptance angle is the numerical aperture of the fiber. These two parameters of the fiber provide a guideline for efficiently coupling light from free space into the fiber core, which is one of the most common practices in optical fiber communication systems.

The same demonstration can be also employed to assist the teacher to explain to the students a very important loss mechanism of the optical fiber: Rayleigh scattering. Generally speaking, Rayleigh scattering refers to the scattering of light off of particles that are small compared to the wavelength of the light, and it accounts for about 96% of attenuation in the optical fiber. What causes Rayleigh scattering in the optical fiber? It is the impurities that cannot be completely eliminated during the manufacturing process. When the light travels in the glass fiber core, it collides with the impurity atoms in the glass and, as a result, the light is scattered. Rayleigh scattering can be considered to be elastic scattering since the wavelength of the scattered light is not changed. If the scattered light maintains an angle that supports forward travel within the fiber core (in our demonstration, the glass rod), no attenuation occurs. If the light is scattered at an angle that does not support continued forward travel, the light is diverted out of the core and attenuation occurs. Here is a good question for the teacher to toss out to the students before the lecture: why can we see the light path sideways, although the light travels in the forward direction in the glass rod? Since we can see the light rays sideways, there must be some light leaving the glass rod through the side and reaching our eyes. But what causes this portion of the light to deviate from its original propagation direction and to exit the glass rod sooner than expected? As the students are now eager to know the answer, the timely introduction of Rayleigh scattering aided with the visual demonstration by the teacher will leave a deep impression in students' minds.

For no-prior-experience-needed lab experiments, students can use an optical signal path demonstrator set [9] to observe the propagation of light in fiber optics. This set includes a straight and a curved acrylic bar. The entire path of a beam of light is visible inside the specially prepared plastic, which contains tiny particles that scatter the light. Both bars work with a classroom laser. Moreover, the students can learn fiber optics by experimenting and building projects using a fiber optics kit, for example [9]. This fiber optical kit includes all fiber optic components, connectors and cables. The students can get acquainted with the light pipe, the fiber optic receiver and the fiber optic light- pen cable.

Teaching can be effective and efficient only when one has the students' attention. The demonstrations described in this paper illustrate our approach to active involvement of the students into the fiber optics lecture and course. We are currently developing methods in the format of laboratory hand-on projects, academic seminars, student group projects, etc., to foster a dynamic teaching and learning environment for both the faculty and the students. The goal of our effort is for the students to engage productively in their learning and to succeed in the course.

## 4. Conclusion

We have briefly reviewed the history and the current development of fiber optics, and have addressed the significance of teaching fiber optics in college for science and non-science majors. Furthermore, we have proposed several simple and illustrative experimental demonstrations for college faculties to use in their teaching in fiber optics. These demonstrations are designed to aid the teaching and learning processes in the lectures. More methods serving the same purpose but in various format including the laboratory projects are currently being developed in our college [11].